\documentclass[printer]{tMOP2e}

\begin{document}

 \doi{10.1080/0950034YYxxxxxxxx}

 \newcommand{\bra}[1]{\langle #1|}
 \newcommand{\ket}[1]{| #1\rangle}
 \newcommand{\braket}[2]{\langle #1|#2\rangle}

\markboth{Jaeger and Ann}{Disentanglement and decoherence in a
pair of qutrits under dephasing noise\\}

\title{Disentanglement and decoherence in a pair of
qutrits under dephasing noise\\}

\author{Gregg Jaeger\thanks{$^\ast$Corresponding
author. Email: jaeger@bu.edu}$^{\ast 1}$ and
Kevin Ann\thanks{$^\dagger$Email: kevinann@bu.edu}$^{\dagger 2}$
\\ \vspace{6pt}
$^{1}$Quantum Imaging Lab, Dept. of Electrical and Computer
Engineering, and Div. of Natural Sciences, Boston University,
Boston, MA 02215
\\ \vspace{6pt}
$^{2}$Physics Department, Boston University, Boston, MA 02215
\\ \vspace{6pt}
\received{v1.0 released April 2006}
}
\maketitle

\begin{abstract}
We relate disentanglement and decoherence rates in a pair of
three-level atoms subjected to multi-local and collective pure
dephasing noise acting in a preferred basis. The bipartite
entanglement decay rate, as bounded from above by the negativity, is
found to be greater than or equal to the dephasing-decoherence rates
characterized by the decay of off-diagonal elements in the
corresponding full density matrix describing the system or the
reduced density matrix describing either qutrit, extending previous
results for qubit pairs subject to such noise.
\end{abstract}

\section{Introduction}
Quantum entanglement can be affected by a range of noise sources,
both quantum and classical in nature. Noise can give rise to a loss
of entanglement over a broad range of quantum states. Recently, the
relationship between dephasing-decoherence and bipartite
entanglement reduction under basis-specific classical noise has been
studied in two-qubit systems
\cite{YE02,YE03,YE04,YE06,TP05,STP06,AY07} and in pairs of qutrit
systems \cite{FS06,DJ06,CM00}; this relationship has been studied in
two-qubit systems under quantum dissipative vacuum noise as well
\cite{CW06}. For example, it has been noted that initially entangled
two-qubit systems can suffer ``entanglement sudden death,'' in which
a two-qubit system may suddenly lose entanglement in a finite time,
even though each qubit itself maintains its
quantum-coherence\cite{YE04,YE06}. Here, we present the first
general analysis of the disentanglement and dephasing of two qutrits
as realized in atoms with ``V''-type energy-level configuration
under a classical pure dephasing noise model, at the multi-local and
collective level. We compare the timescales of disentanglement and
dephasing-decoherence, the latter timescale specifically in the
basis on which this noise acts as in previous studies.

In Sec. 2, we introduce our dephasing model.  In Sec. 3, we examine
the effects of the multi-local and collective dephasing noise on a
general state.  Finally, in Sec. 4, we examine two specific classes
of states and compare their disentanglement and decoherence rates
explicitly. The bipartite-entanglement decay rate, is found to be
greater than or equal to the dephasing-decoherence rates,
characterized by the decay of off-diagonal elements in the
corresponding density matrix describing the full system or the
reduced density matrix describing either qutrit, when dephasing
occurs. This result is the most general yet to be obtained in the
comparative study of decoherence and disentanglement in two-qutrit
systems.

\section{Model}
Our model consists of two-three level systems subjected to
the time-dependent Hamiltonian
\begin{eqnarray}
\tiny
H(t)=
-\frac{\mu}{2}
\left[
b^{(1)}_{{\rm A}}\left(t\right)
\sigma_{\rm z}^{\rm A}
+
b^{(1)}_{{\rm B}}\left(t\right)
\sigma_{\rm z}^{\rm B}
+
b^{(2)}_{{\rm AB}}\left(t\right)
\left(
\sigma_{\rm z}^{\rm A}
+
\sigma_{\rm z}^{\rm B}
\right)
\right],
\end{eqnarray}
\noindent where $\sigma_{\rm z} = {\rm diag}(1, e^{\frac{i 2
\pi}{3}}, e^{\frac{i 4 \pi}{3}})$ is the dephasing operator for
three-level systems with subscripts denoting qutrits A, B, or both,
$\hbar = 1$, the time dependent noise terms $b_X^{(i)}(i=1,2)$ refer
to statistically independent classical Markov processes satisfying
$\left\langle b_{X}\left(t\right) \right\rangle = 0 \ {\rm and \ }
\left\langle b_{X}\left(t\right) b_{X}\left(t'\right) \right\rangle
= \frac{\Gamma_{1}}{\mu^2}\delta\left(t-t'\right),$ with $X = {\rm
A},{\rm B}$; $\left\langle b_{\rm AB}\left(t\right) \right\rangle =
0 \ {\rm and \ } \left\langle b_{\rm AB}\left(t\right) b_{\rm
AB}\left(t'\right) \right\rangle =
\frac{\Gamma_{2}}{\mu^2}\delta\left(t-t'\right)$; $\left\langle
\cdots \right\rangle$ is the ensemble time average, $\Gamma_{\rm 1}$
and $\Gamma_{\rm 2}$ denote the phase-damping rates associated with
$b_{X}(t)$ (X = A, B) and $b_{\rm AB}\left(t\right)$, respectively.

The one-qutrit standard-basis eigenstates are $\left\{\ket{0},
\ket{+1}, \ket{-1}\right\}\label{1QutritBasis}$, for example,
representing the ground state, first-excited state, and
second-excited state of the atom, respectively. We assume that the
states $|+1\rangle ,|-1\rangle$ couple to the ground state but not
to each other. Here, we notate the standard two-qutrit basis states
via the obvious one-to-one correspondence $\left\{\ket{1}, \ket{2},
\ket{3}, \ket{4}, \ket{5}, \ket{6}, \ket{7}, \ket{8},
\ket{9}\right\} \doteq\left\{ \ket{00}, \ket{0,+1}, \ket{0,-1},
\ket{+1,0}, \ket{+1,+1}, \ket{+1,-1}, \ket{-1,0}, \ket{-1,+1},
\ket{-1,-1} \right\} \label{2QutritBasis}$, for simplicity.

The time-dependent density matrix for the two-qutrit system is
obtained by taking ensemble averages over the three noise fields,
$b_{{\rm A}}\left(t\right)$, $b_{{\rm B}}\left(t\right)$, $b_{{\rm
AB}}\left(t\right)$, that is, $ \rho\left(t\right) = \left\langle
\rho_{\rm st}\left(t\right) \right\rangle_{{\rm A (B, AB)}},
\label{rhoAverage} $ where the statistical density operator
$\rho_{\rm st}\left(t\right)$ and the unitary operator
$U\left(t\right)$ associated with $H(t)$ are $ \rho_{\rm
st}\left(t\right) = U\left(t\right)\rho\left(0\right)
U^{\dagger}\left(t\right) {\rm and} \ U\left(t\right) =
\exp\left[-i\int_{0}^{t}{dt' H\left(t'\right)}\right]\ , $
respectively. It is helpful to consider the dynamical evolution of
$\rho(t)$ as a completely positive trace preserving (CPTP) linear
map $\mathcal{E}(\rho)$, that is, a combination of local and
collective quantum channels, any of which can be turned off in
particular cases, taking an input state $\rho\left(0\right)$ to the
output state $\rho\left(t\right)$ given by the operator sum
$\rho\left(t\right) = \mathcal{E}\left(\rho\left(0\right)\right) =
\sum_{\mu =
1}^{N}\overline{E}_{\mu}^{\dagger}\left(t\right)\rho\left(0\right)
\overline{E}_{\mu}\left(t\right)\label{kraussSumDefinition},$ where
$\overline{E}_{\mu}$ are decomposition operators that satisfy the
completeness relation $
\sum_{\mu}\overline{E}_{\mu}^{\dagger}\overline{E}_{\mu} =
\mathbb{I}\ . $ In each of the various cases considered here, the
internal structure of the $\overline{E}_{\mu}$ in accordance with
the Hamiltonian; various terms may or may not nontrivially
contribute in a given case. The most general solution of
$\rho\left(t\right)$, assuming that the system is not initially
correlated with any of the three environments, is
$\rho\left(t\right) = \sum_{i,j=1}^{3}\sum_{k=1}^{3}
\big(D_{k}^{{\rm AB}\dagger}E_{j}^{{\rm B}\dagger}E_{i}^{{\rm
A}\dagger} \big) \rho\left(0\right) \left(E_{i}^{\rm A}E_{j}^{\rm
B}D_{k}^{\rm AB}\right)$, where the terms describing the interaction
with the local magnetic fields $b_{{\rm A}}\left(t\right)$ and
$b_{{\rm B}}\left(t\right)$ involve the decomposition operators $
E_{1}^{{\rm A}} = {\rm diag}( 1, \gamma_{\rm A}\left(t\right),
\gamma_{\rm A}\left(t\right)) \otimes \mathbb{I}_{3}, 
E_{2}^{{\rm
A}} = {\rm diag}(0, \omega_{\rm A}\left(t\right), 0) \otimes \mathbb{I}_{3}, 
E_{3}^{{\rm
A}} = {\rm diag}(0, 0, \omega_{\rm
A}\left(t\right)) \otimes \mathbb{I}_{3}, 
E_{1}^{{\rm B}} =
\mathbb{I}_{3} \otimes {\rm diag}(1, \gamma_{\rm B}\left(t\right),
\gamma_{\rm B}\left(t\right)), 
E_{2}^{{\rm B}} =
\mathbb{I}_{3} \otimes {\rm diag}(0, \omega_{\rm B}\left(t\right), 0),
{\rm and} \ E_{2}^{{\rm B}} =
\mathbb{I}_{3} \otimes {\rm diag}(0, 0,
\omega_{\rm B}\left(t\right)),$ and the terms associated with the
global magnetic field $b_{{\rm AB}}\left(t\right)$ involve the
operators $D_{1}^{\rm AB} = {\rm diag}(\gamma_{\rm
AB}\left(t\right),1,1,1, \gamma_{\rm
AB}\left(t\right),1,1,1,\gamma_{\rm AB}\left(t\right)), \ $
$D_{2}^{\rm AB} = {\rm diag}(\omega_{\rm AB1}\left(t\right),0,0,0,
\omega_{\rm AB2}\left(t\right),0,0,0,\omega_{\rm
AB2}\left(t\right)), \ $ $D_{3}^{\rm AB} = {\rm
diag}(0,0,0,0,\omega_{\rm AB3}\left(t\right),0,0,0,\omega_{\rm AB3}\left(t\right)) \ .$
%
The time-dependent parameters appearing in the matrices above are
given by: $ \gamma_{i}\left(t\right) = e^{-t/2T_{i}},
\omega_{i}\left(t\right) = \sqrt{1-\gamma_{i}^{2}},
\gamma\left(t\right) = e^{-t/2T_{i}}, \omega_{i 1}\left(t\right) =
\sqrt{1-\gamma_{i}^{2}}, \omega_{i 2}\left(t\right) =
-\gamma_{i}^{2}\sqrt{1-\gamma_{i}^{2}},{\rm and} \ \omega_{i
3}\left(t\right) = \sqrt{
\left(1-\gamma_{i}^{2}\right)\left(1-\gamma_{i}^{4} \right) }, $
where $T_{i}=\frac{1}{\Gamma_{i}} (i=1,2)$ are the phase-relaxation
times associated with the the pertinent qubits ${\rm A}$ and ${\rm
B}$, respectively, as introduced in \cite{YE03}, with $\Gamma_{i}$
being the rate parameters.  From here on, for tractability of
notation, time does not explicitly appear as an argument for these
quantities but is implied.

Dephasing-decoherence rates are characterized by the decay of
off-diagonal elements in the corresponding full density matrix
describing the system or the reduced density matrix describing
either qutrit given via characteristic decay times. Entanglement is
bounded from above by the negativity $ \mathcal{N}\left(\rho\right)
= \frac{\left\| \rho^{{\rm T}_{\rm A}} \right\|_{1} - 1}{2},
\label{negativity} $ where $\rho^{{\rm T}_{\rm A}}$ is the partial
transpose of the density matrix with respect to qutrit A and
$\left\| \cdot \right\|_{1}$ denotes the trace norm \cite{VW02}.

\section{General Case}
In the standard-basis representation of Eq. \ref{2QutritBasis}, the
generic pure state of the two-qutrit system is $ \scriptsize
\ket{\Psi}_{{\rm AB}} = \bar{a}_{1}\ket{1} + \bar{a}_{2}\ket{2} +
\bar{a}_{3}\ket{3} + \bar{a}_{4}\ket{4} + \bar{a}_{5}\ket{5} +
\bar{a}_{6}\ket{6} + \bar{a}_{7}\ket{7} + \bar{a}_{8}\ket{8} +
\bar{a}_{9}\ket{9}, $ a normalized state-vector with $\bar{a}_{i}
\in \mathbb{C}$ and $\sum^{9}_{i=1}\bar{a}^{2}_{i} = 1$. Our
analysis proceeds as follows.  We find the explicit time evolution
of the general state subjected to noise from the multi-local
dephasing channel $\mathcal{EF}$ and the collective dephasing
channel $\mathcal{D}$. The decoherence timescales are then
determined, as characterized by the decay of the off-diagonal
elements at the level of the full density matrix of two-qutrits, as
well as the reduced density matrix of each individual qutrit, for
each of the dephasing cases, multi-local and collective. We then
analyze disentanglement timescales, using the monotone of
negativity, when there is decoherence for the general state in each
of these cases. In order to compare decoherence and disentanglement
behavior explicitly, we then consider the the behavior of two
particular subclasses of states, the {\it robust} class and the {\it
fragile} class.

First, consider the most general initial two-qutrit pure state,
$\rho_{{\rm AB}}\left(0\right)=P(\ket{\Psi}_{{\rm AB}}),$
where $P(\ket{\Psi}_{{\rm AB}})$ is the projector corresponding to
the generic state-vector argument $\ket{\Psi}_{\rm AB}$ explicitly
given above, under multi-local and collective dephasing noise.

\subsection{General Case: Multi-Local Dephasing Channel $\mathcal{EF}$}
\begin{eqnarray}
\rho^{{\rm G},\mathcal EF}_{\rm AB}\left(t\right) =
\left(
\begin{array}{lllllllll}
\left|\bar{a}_{1}\right|^{2} &
 \bar{a}_{1}\bar{a}_{2}^{\ast}\gamma_{\rm B} &
 \bar{a}_{1}\bar{a}_{3}^{\ast}\gamma_{\rm B} &
 \bar{a}_{1}\bar{a}_{4}^{\ast}\gamma_{\rm A} &
 \bar{a}_{1}\bar{a}_{5}^{\ast}\gamma_{\rm A}\gamma_{\rm B} &
 \bar{a}_{1}\bar{a}_{6}^{\ast}\gamma_{\rm A}\gamma_{\rm B} &
 \bar{a}_{1}\bar{a}_{7}^{\ast}\gamma_{\rm A} &
 \bar{a}_{1}\bar{a}_{8}^{\ast}\gamma_{\rm A}\gamma_{\rm B} &
 \bar{a}_{1}\bar{a}_{9}^{\ast}\gamma_{\rm A}\gamma_{\rm B} \\ 
\bar{a}_{2}\bar{a}_{1}^{\ast}\gamma_{\rm B} &
 \left|\bar{a}_{2}\right|^{2} &
 \bar{a}_{2}\bar{a}_{3}^{\ast}\gamma_{\rm B}^{2} &
 \bar{a}_{2}\bar{a}_{4}^{\ast}\gamma_{\rm A}\gamma_{\rm B} &
 \bar{a}_{2}\bar{a}_{5}^{\ast}\gamma_{\rm A} &
 \bar{a}_{2}\bar{a}_{6}^{\ast}\gamma_{\rm A}\gamma_{\rm B}^{2} &
 \bar{a}_{2}\bar{a}_{7}^{\ast}\gamma_{\rm A}\gamma_{\rm B} &
 \bar{a}_{2}\bar{a}_{8}^{\ast}\gamma_{\rm A} &
 \bar{a}_{2}\bar{a}_{9}^{\ast}\gamma_{\rm A}\gamma_{\rm B}^{2} \\ 
\bar{a}_{3}\bar{a}_{1}^{\ast}\gamma_{\rm B} &
 \bar{a}_{3}\bar{a}_{2}^{\ast}\gamma_{\rm B}^{2} &
 \left|\bar{a}_{3}\right|^{2} &
 \bar{a}_{3}\bar{a}_{4}^{\ast}\gamma_{\rm A}\gamma_{\rm B} &
 \bar{a}_{3}\bar{a}_{5}^{\ast}\gamma_{\rm A}\gamma_{\rm B}^{2} &
 \bar{a}_{3}\bar{a}_{6}^{\ast}\gamma_{\rm A} &
 \bar{a}_{3}\bar{a}_{7}^{\ast}\gamma_{\rm A}\gamma_{\rm B} &
 \bar{a}_{3}\bar{a}_{8}^{\ast}\gamma_{\rm A}\gamma_{\rm B}^{2} &
 \bar{a}_{3}\bar{a}_{9}^{\ast}\gamma_{\rm A} \\ 
\bar{a}_{4}\bar{a}_{1}^{\ast}\gamma_{\rm A} &
 \bar{a}_{4}\bar{a}_{2}^{\ast}\gamma_{\rm A}\gamma_{\rm B} &
 \bar{a}_{4}\bar{a}_{3}^{\ast}\gamma_{\rm A}\gamma_{\rm B} &
 \left|\bar{a}_{4}\right|^{2} &
 \bar{a}_{4}\bar{a}_{5}^{\ast}\gamma_{\rm B} &
 \bar{a}_{4}\bar{a}_{6}^{\ast}\gamma_{\rm B} &
 \bar{a}_{4}\bar{a}_{7}^{\ast}\gamma_{\rm A}^{2} &
 \bar{a}_{4}\bar{a}_{8}^{\ast}\gamma_{\rm A}^{2}\gamma_{\rm B} &
 \bar{a}_{4}\bar{a}_{9}^{\ast}\gamma_{\rm A}^{2}\gamma_{\rm B} \\ 
\bar{a}_{5}\bar{a}_{1}^{\ast}\gamma_{\rm A}\gamma_{\rm B} &
 \bar{a}_{5}\bar{a}_{2}^{\ast}\gamma_{\rm A} &
 \bar{a}_{5}\bar{a}_{3}^{\ast}\gamma_{\rm A}\gamma_{\rm B}^{2} &
 \bar{a}_{5}\bar{a}_{4}^{\ast}\gamma_{\rm B} &
 \left|\bar{a}_{5}\right|^{2} &
 \bar{a}_{5}\bar{a}_{6}^{\ast}\gamma_{\rm B}^{2} &
 \bar{a}_{5}\bar{a}_{7}^{\ast}\gamma_{\rm A}^{2}\gamma_{\rm B} &
 \bar{a}_{5}\bar{a}_{8}^{\ast}\gamma_{\rm A}^{2} &
 \bar{a}_{5}\bar{a}_{9}^{\ast}\gamma_{\rm A}^{2}\gamma_{\rm B}^{2} \\ 
\bar{a}_{6}\bar{a}_{1}^{\ast}\gamma_{\rm A}\gamma_{\rm B} &
 \bar{a}_{6}\bar{a}_{2}^{\ast}\gamma_{\rm A}\gamma_{\rm B}^{2} &
 \bar{a}_{6}\bar{a}_{3}^{\ast}\gamma_{\rm A} &
 \bar{a}_{6}\bar{a}_{4}^{\ast}\gamma_{\rm B} &
 \bar{a}_{6}\bar{a}_{5}^{\ast}\gamma_{\rm B}^{2}&
 \left|\bar{a}_{6}\right|^{2} &
 \bar{a}_{6}\bar{a}_{7}^{\ast}\gamma_{\rm A}^{2}\gamma_{\rm B} &
 \bar{a}_{6}\bar{a}_{8}^{\ast}\gamma_{\rm A}^{2}\gamma_{\rm B}^{2} &
 \bar{a}_{6}\bar{a}_{9}^{\ast}\gamma_{\rm A}^{2} \\ 
\bar{a}_{7}\bar{a}_{1}^{\ast}\gamma_{\rm A} &
 \bar{a}_{7}\bar{a}_{2}^{\ast}\gamma_{\rm A}\gamma_{\rm B} &
 \bar{a}_{7}\bar{a}_{3}^{\ast}\gamma_{\rm A}\gamma_{\rm B} &
 \bar{a}_{7}\bar{a}_{4}^{\ast}\gamma_{\rm A}^{2} &
 \bar{a}_{7}\bar{a}_{5}^{\ast}\gamma_{\rm A}^{2}\gamma_{\rm B} &
 \bar{a}_{7}\bar{a}_{6}^{\ast}\gamma_{\rm A}^{2}\gamma_{\rm B} &
 \left|\bar{a}_{7}\right|^{2} &
 \bar{a}_{7}\bar{a}_{8}^{\ast}\gamma_{\rm B} &
 \bar{a}_{7}\bar{a}_{9}^{\ast}\gamma_{\rm B} \\ 
\bar{a}_{8}\bar{a}_{1}^{\ast}\gamma_{\rm A}\gamma_{\rm B} &
 \bar{a}_{8}\bar{a}_{2}^{\ast}\gamma_{\rm A} &
 \bar{a}_{8}\bar{a}_{3}^{\ast}\gamma_{\rm A}\gamma_{\rm B}^{2} &
 \bar{a}_{8}\bar{a}_{4}^{\ast}\gamma_{\rm A}^{2}\gamma_{\rm B} &
 \bar{a}_{8}\bar{a}_{5}^{\ast}\gamma_{\rm A}^{2} &
 \bar{a}_{8}\bar{a}_{6}^{\ast}\gamma_{\rm A}^{2}\gamma_{\rm B}^{2} &
 \bar{a}_{8}\bar{a}_{7}^{\ast}\gamma_{\rm B} &
 \left|\bar{a}_{8}\right|^{2} &
 \bar{a}_{8}\bar{a}_{9}^{\ast}\gamma_{\rm B}^{2} \\ 
\bar{a}_{9}\bar{a}_{1}^{\ast}\gamma_{\rm A}\gamma_{\rm B} &
 \bar{a}_{9}\bar{a}_{2}^{\ast}\gamma_{\rm A}\gamma_{\rm B}^{2} &
 \bar{a}_{9}\bar{a}_{3}^{\ast}\gamma_{\rm A} &
 \bar{a}_{9}\bar{a}_{4}^{\ast}\gamma_{\rm A}^{2}\gamma_{\rm B} &
 \bar{a}_{9}\bar{a}_{5}^{\ast}\gamma_{\rm A}^{2}\gamma_{\rm B}^{2} &
 \bar{a}_{9}\bar{a}_{6}^{\ast}\gamma_{\rm A}^{2} &
 \bar{a}_{9}\bar{a}_{7}^{\ast}\gamma_{\rm B} &
 \bar{a}_{9}\bar{a}_{8}^{\ast}\gamma_{\rm B}^{2} &
 \left|\bar{a}_{9}\right|^{2} \\ 
\end{array}
\right)
\end{eqnarray}
is the time-evolved full density matrix for multi-local dephasing.
Our density matrices are labeled from here on as $\rho_{\rm X}^{{\rm
c},\mathcal{C}}$, with X denoting the pertinent qubits (A, B, AB), c
is the class examined (G,F,R) as described in the next section, and
$\mathcal{C}$ denoting the pertinent dephasing channel ($\mathcal{E,
F, D}$ ) As noted above, for tractability of notation, time does not
explicitly appear as an argument for the $\gamma_{A}$ and
$\gamma_{B}$.  Note that one can recover the local dephasing channel
$\mathcal{E}(\mathcal{F})$ for qutrit A(B) by effectively freezing
the time parameter in the exponential of the
$\gamma_{B}$($\gamma_{A}$) factors of the opposite channel. The
effect of two local dephasing noise sources, one for each qutrit, is
simply the combination of independent effects of the individual
local dephasing channel just described. Thus, the combined matrix is
just the component-wise multiplication of exponential decay factors
arising from each of the local dephasing channels. Due to their
appearance in exponents, the decay rates add.

Let us designate the different timescales as
$\tau_{i{\rm-dec},{\mathcal{C}}}^{c(j)} \ \ {\rm and} \ \
\tau_{i{\rm-dis},{\mathcal{C}}}^{c(j)} $, representing the
decoherence and disentanglement timescales, respectively; $i$
denotes the number of qutrits affected ($i=1,2$), $\mathcal{C}$
denotes the pertinent dephasing channel ($\mathcal{E,F,D}$), $j$ is
used as an index further to discriminate the timescales, and c
denotes the class examined (G,F,R). The differing timescales of
reduction of off-diagonal elements consist of various combinations
of $\gamma_{A}$ and $\gamma_{B}$. Here, we have the four timescales
of decay as $ \tau_{\rm 2-dec,\mathcal{EF}}^{G(1)} =
2\big({1\over\Gamma_{\rm 1}}\big), \ \tau_{\rm
2-dec,\mathcal{EF}}^{G(2)} = \big({1\over\Gamma_{\rm 1}}\big), \
\tau_{\rm 2-dec,\mathcal{EF}}^{G(3)} = 2\big({1\over3\Gamma_{\rm
1}}\big), \ {\rm and} \ \tau_{\rm 2-dec,\mathcal{EF}}^{G(4)} =
\big({1\over 2\Gamma_{\rm 1}}\big). $
The reduced density matrices of the individual qutrit subsystems are
\begin{eqnarray}
\rho^{{\rm G},\mathcal EF}_{{\rm A}}\left(t\right)=
{\rm Tr}_{\rm B}\rho^{{\rm G},\mathcal EF}_{{\rm AB}}\left(t\right) =
\left(
\begin{array}{lll}
 \left|\bar{a}_{1}\right|^{2} +
 \left|\bar{a}_{2}\right|^{2} +
 \left|\bar{a}_{3}\right|^{2}
 &
\left(
 \bar{a}_{1}\bar{a}_{4}^{\ast} +
 \bar{a}_{2}\bar{a}_{5}^{\ast} +
 \bar{a}_{3}\bar{a}_{6}^{\ast}
 \right)\gamma_{\rm A}
 &
\left(
 \bar{a}_{1}\bar{a}_{7}^{\ast} +
 \bar{a}_{2}\bar{a}_{8}^{\ast} +
 \bar{a}_{3}\bar{a}_{9}^{\ast}
\right)\gamma_{\rm A}, \\
\left(
 \bar{a}_{4}\bar{a}_{1}^{\ast} +
 \bar{a}_{5}\bar{a}_{2}^{\ast} +
 \bar{a}_{6}\bar{a}_{3}^{\ast}
 \right)\gamma_{\rm A}
 &
 \left|\bar{a}_{4}\right|^{2} +
 \left|\bar{a}_{5}\right|^{2} +
 \left|\bar{a}_{6}\right|^{2}
 &
\left(
 \bar{a}_{4}\bar{a}_{7}^{\ast} +
 \bar{a}_{5}\bar{a}_{8}^{\ast} +
 \bar{a}_{6}\bar{a}_{9}^{\ast}
 \right)\gamma_{\rm A}^{2} \\
\left(
 \bar{a}_{7}\bar{a}_{1}^{\ast} +
 \bar{a}_{8}\bar{a}_{2}^{\ast} +
 \bar{a}_{9}\bar{a}_{3}^{\ast}
 \right)\gamma_{\rm A}
 &
\left(
 \bar{a}_{7}\bar{a}_{4}^{\ast} +
 \bar{a}_{8}\bar{a}_{5}^{\ast} +
 \bar{a}_{9}\bar{a}_{6}^{\ast}
 \right)\gamma_{\rm A}^{2}
 &
\left|\bar{a}_{7}\right|^{2} +
 \left|\bar{a}_{8}\right|^{2} +
 \left|\bar{a}_{9}\right|^{2}
 \\
\end{array}
\right)\\
\rho^{{\rm G},\mathcal EF}_{{\rm B}}\left(t\right)=
{\rm Tr}_{\rm A}\rho^{{\rm G},\mathcal EF}_{{\rm AB}}\left(t\right) =
\left(
\begin{array}{lll}
 \left|\bar{a}_{1}\right|^{2} +
 \left|\bar{a}_{4}\right|^{2} +
 \left|\bar{a}_{7}\right|^{2}
 &
\left(
 \bar{a}_{1}\bar{a}_{2}^{\ast} +
 \bar{a}_{4}\bar{a}_{5}^{\ast} +
 \bar{a}_{7}\bar{a}_{8}^{\ast}
 \right)\gamma_{\rm B}
 &
\left(
 \bar{a}_{1}\bar{a}_{3}^{\ast} +
 \bar{a}_{4}\bar{a}_{6}^{\ast} +
 \bar{a}_{7}\bar{a}_{9}^{\ast}
\right)\gamma_{\rm B} \\
\left(
 \bar{a}_{2}\bar{a}_{1}^{\ast} +
 \bar{a}_{5}\bar{a}_{4}^{\ast} +
 \bar{a}_{8}\bar{a}_{7}^{\ast}
 \right)\gamma_{\rm B}
 &
 \left|\bar{a}_{2}\right|^{2} +
 \left|\bar{a}_{5}\right|^{2} +
 \left|\bar{a}_{8}\right|^{2}
 &
\left(
 \bar{a}_{2}\bar{a}_{3}^{\ast} +
 \bar{a}_{5}\bar{a}_{5}^{\ast} +
 \bar{a}_{8}\bar{a}_{9}^{\ast}
 \right)\gamma_{\rm B}^{2} \\
\left(
 \bar{a}_{3}\bar{a}_{1}^{\ast} +
 \bar{a}_{6}\bar{a}_{4}^{\ast} +
 \bar{a}_{9}\bar{a}_{7}^{\ast}
 \right)\gamma_{\rm B}
 &
\left(
 \bar{a}_{3}\bar{a}_{2}^{\ast} +
 \bar{a}_{6}\bar{a}_{5}^{\ast} +
 \bar{a}_{9}\bar{a}_{8}^{\ast}
 \right)\gamma_{\rm B}^{2}
 &
\left|\bar{a}_{3}\right|^{2} +
 \left|\bar{a}_{6}\right|^{2} +
 \left|\bar{a}_{9}\right|^{2}
 \\
\end{array}
\right).\
\end{eqnarray}
Again, one finds differing dephasing decoherence timescales. In
particular, the off-diagonal elements decay in two different
timescales before the density matrix becomes fully diagonal in the
basis under consideration, $ \tau_{\rm 1-dec,\mathcal{EF}}^{G(1)} =
2\big({1\over\Gamma_{\rm 1}}\big) \ {\rm and} \ \tau_{\rm
1-dec,\mathcal{EF}}^{G(2)} = \big({1\over\Gamma_{\rm 1}}\big) . $

The negativity, which bounds the (non-bound) entanglement,
\begin{eqnarray}
\tiny \mathcal{N}\left[(\rho_{\rm
AB}^{\mathcal{EF}}\left(t\right))^{\rm T_{A}}\right] &=&
\frac{1}{2}\bigg(-1 + \sqrt{ \left[ \left|\bar{a}_{1}\right|^{2} +
(\left|\bar{a}_{4}\right|^{2} +
\left|\bar{a}_{7}\right|^{2})\gamma_{\rm A}^{2} \right] \left[
\left|\bar{a}_{1}\right|^{2} + (\left|\bar{a}_{2}\right|^{2} +
\left|\bar{a}_{3}\right|^{2})\gamma_{\rm B}^{2} \right]
} \nonumber \\ 
&+& \sqrt{
\left[
\left|\bar{a}_{7}\right|^{2} + \left|\bar{a}_{1}\right|^{2}\gamma_{\rm A}^{2} +
\left|\bar{a}_{4}\right|^{2}\gamma_{\rm A}^{4}
\right]
\left[
\left|\bar{a}_{7}\right|^{2} + (\left|\bar{a}_{8}\right|^{2} +
\left|\bar{a}_{9}\right|^{2})\gamma_{\rm B}^{2}
\right]
} \nonumber \\ 
&+& \sqrt{
\left[
\left|\bar{a}_{4}\right|^{2} + \left|\bar{a}_{1}\right|^{2}\gamma_{\rm A}^{2} +
\left|\bar{a}_{7}\right|^{2}\gamma_{\rm A}^{4}
\right]
\left[
\left|\bar{a}_{4}\right|^{2} + (\left|\bar{a}_{5}\right|^{2} +
\left|\bar{a}_{6}\right|^{2})\gamma_{\rm B}^{2}
\right]
} \nonumber \\ 
&+& \sqrt{
\left[
\left|\bar{a}_{3}\right|^{2} + (\left|\bar{a}_{6}\right|^{2} +
\left|\bar{a}_{9}\right|^{2})\gamma_{\rm A}^{2}
\right]
\left[
\left|\bar{a}_{3}\right|^{2} + \left|\bar{a}_{1}\right|^{2}\gamma_{\rm B}^{2} +
\left|\bar{a}_{2}\right|^{2}\gamma_{\rm B}^{4}
\right]
} \nonumber \\ 
&+& \sqrt{
\left[
\left|\bar{a}_{6}\right|^{2} + \left|\bar{a}_{3}\right|^{2}\gamma_{\rm A}^{2} +
\left|\bar{a}_{9}\right|^{2}\gamma_{\rm A}^{4}
\right]
\left[
\left|\bar{a}_{6}\right|^{2} + \left|\bar{a}_{4}\right|^{2}\gamma_{\rm B}^{2} +
\left|\bar{a}_{5}\right|^{2}\gamma_{\rm B}^{4}
\right]
} \nonumber \\ 
&+& \sqrt{
\left[
\left|\bar{a}_{9}\right|^{2} + \left|\bar{a}_{3}\right|^{2}\gamma_{\rm A}^{2} +
\left|\bar{a}_{6}\right|^{2}\gamma_{\rm A}^{4}
\right]
\left[
\left|\bar{a}_{9}\right|^{2} + \left|\bar{a}_{7}\right|^{2}\gamma_{\rm B}^{2} +
\left|\bar{a}_{8}\right|^{2}\gamma_{\rm B}^{4}
\right]
} \nonumber \\ 
&+& \sqrt{
\left[
\left|\bar{a}_{2}\right|^{2} + (\left|\bar{a}_{5}\right|^{2} +
\left|\bar{a}_{8}\right|^{2})\gamma_{\rm A}^{2}
\right]
\left[
\left|\bar{a}_{2}\right|^{2} + \left|\bar{a}_{1}\right|^{2}\gamma_{\rm B}^{2} +
\left|\bar{a}_{3}\right|^{2}\gamma_{\rm B}^{4}
\right]
} \nonumber \\ 
&+& \sqrt{
\left[
\left|\bar{a}_{8}\right|^{2} + \left|\bar{a}_{2}\right|^{2}\gamma_{\rm A}^{2} +
\left|\bar{a}_{5}\right|^{2}\gamma_{\rm A}^{4}
\right]
\left[
\left|\bar{a}_{8}\right|^{2} + \left|\bar{a}_{7}\right|^{2}\gamma_{\rm B}^{2} +
\left|\bar{a}_{9}\right|^{2}\gamma_{\rm B}^{4}
\right]
} \nonumber \\ 
&+& \sqrt{ \left[ \left|\bar{a}_{5}\right|^{2} +
\left|\bar{a}_{2}\right|^{2}\gamma_{\rm A}^{2} +
\left|\bar{a}_{8}\right|^{2}\gamma_{\rm A}^{4} \right] \left[
\left|\bar{a}_{5}\right|^{2} +
\left|\bar{a}_{4}\right|^{2}\gamma_{\rm B}^{2} +
\left|\bar{a}_{6}\right|^{2}\gamma_{\rm B}^{4} \right]}\
\bigg) \ ,
\end{eqnarray}
is rather unwieldy. It is therefore illuminating to characterize its
behavior for each of the dephasing cases: no dephasing, one-qutrit
local dephasing, and two-qutrit multi-local dephasing. In the
expression for negativity, we can isolate the behavior of one local
dephasing channel with respect to the other by setting the decay
factors corresponding to the other channel to one, effectively
freezing that factor to time zero.

\noindent (1) In cases where there is no dephasing, any decay
factors multiplying probability amplitudes do not fall off
exponentially but instead their value remains at unity.
$\mathcal{N}\left(\rho\left(0\right)\right) =
\mathcal{N}\left(\rho\left(t\right)\right) = 1$, which corresponds
to a maximally entangled state for all time. Thus the (trivial)
dephasing and decoherence rates are equal.

\noindent (2) In the case of a single local dephasing channel,
either channel $\mathcal{E}$ on qutrit A only or channel
$\mathcal{F}$ on qutrit B only, the  negativity
$\mathcal{N}\left(\rho\left(t\right)\right)\rightarrow {\rm c} \ge
0$ in the large-time limit.

\noindent (3) Finally, in the case in which there is dephasing on
both local channels $\mathcal{E}$ and $\mathcal{F}$,
$\mathcal{N}\left(\rho\left(t\right)\right)\rightarrow 0$ in the
large-time limit.

One then notes that in all these cases decoherence never proceeds
faster than disentanglement, as we previously found to be the case
for qubit pairs \cite{AY07}. In later sections, when examining
specific classes of states, we compare decoherence and
disentanglement more explicitly.

\subsection{General Case: Collective Dephasing Channel $\mathcal{D}$}
When subjected to collective dephasing noise, the time-evolved
density matrix of the two-qutrit system in general case is given by
\begin{eqnarray}
\rho^{{\rm G},{\mathcal D}}_{{\rm AB}}\left(t\right) =
\left(
\begin{array}{lllllllll}
\left|\bar{a}_{1}\right|^{2} &
 \bar{a}_{1}\bar{a}_{2}^{\ast}\gamma &
 \bar{a}_{1}\bar{a}_{3}^{\ast}\gamma &
 \bar{a}_{1}\bar{a}_{4}^{\ast}\gamma &
 \bar{a}_{1}\bar{a}_{5}^{\ast}\gamma^4 &
 \bar{a}_{1}\bar{a}_{6}^{\ast}\gamma &
 \bar{a}_{1}\bar{a}_{7}^{\ast}\gamma &
 \bar{a}_{1}\bar{a}_{8}^{\ast}\gamma &
 \bar{a}_{1}\bar{a}_{9}^{\ast}\gamma^4 \\ 
\bar{a}_{2}\bar{a}_{1}^{\ast}\gamma &
 \left|\bar{a}_{2}\right|^{2} &
 \bar{a}_{2}\bar{a}_{3}^{\ast} &
 \bar{a}_{2}\bar{a}_{4}^{\ast} &
 \bar{a}_{2}\bar{a}_{5}^{\ast}\gamma &
 \bar{a}_{2}\bar{a}_{6}^{\ast} &
 \bar{a}_{2}\bar{a}_{7}^{\ast} &
 \bar{a}_{2}\bar{a}_{8}^{\ast} &
 \bar{a}_{2}\bar{a}_{9}^{\ast}\gamma \\ 
\bar{a}_{3}\bar{a}_{1}^{\ast}\gamma &
 \bar{a}_{3}\bar{a}_{2}^{\ast} &
 \left|\bar{a}_{3}\right|^{2} &
 \bar{a}_{3}\bar{a}_{4}^{\ast} &
 \bar{a}_{3}\bar{a}_{5}^{\ast}\gamma &
 \bar{a}_{3}\bar{a}_{6}^{\ast} &
 \bar{a}_{3}\bar{a}_{7}^{\ast} &
 \bar{a}_{3}\bar{a}_{8}^{\ast} &
 \bar{a}_{3}\bar{a}_{9}^{\ast}\gamma \\ 
\bar{a}_{4}\bar{a}_{1}^{\ast}\gamma &
 \bar{a}_{4}\bar{a}_{2}^{\ast} &
 \bar{a}_{4}\bar{a}_{3}^{\ast} &
 \left|\bar{a}_{4}\right|^{2} &
 \bar{a}_{4}\bar{a}_{5}^{\ast}\gamma &
 \bar{a}_{4}\bar{a}_{6}^{\ast} &
 \bar{a}_{4}\bar{a}_{7}^{\ast} &
 \bar{a}_{4}\bar{a}_{8}^{\ast} &
 \bar{a}_{4}\bar{a}_{9}^{\ast}\gamma \\ 
\bar{a}_{5}\bar{a}_{1}^{\ast}\gamma^4 &
 \bar{a}_{5}\bar{a}_{2}^{\ast}\gamma &
 \bar{a}_{5}\bar{a}_{3}^{\ast}\gamma &
 \bar{a}_{5}\bar{a}_{4}^{\ast}\gamma &
 \left|\bar{a}_{5}\right|^{2} &
 \bar{a}_{5}\bar{a}_{6}^{\ast}\gamma &
 \bar{a}_{5}\bar{a}_{7}^{\ast}\gamma &
 \bar{a}_{5}\bar{a}_{8}^{\ast}\gamma &
 \bar{a}_{5}\bar{a}_{9}^{\ast} \\ 
\bar{a}_{6}\bar{a}_{1}^{\ast}\gamma &
 \bar{a}_{6}\bar{a}_{2}^{\ast} &
 \bar{a}_{6}\bar{a}_{3}^{\ast} &
 \bar{a}_{6}\bar{a}_{4}^{\ast} &
 \bar{a}_{6}\bar{a}_{5}^{\ast}\gamma&
 \left|\bar{a}_{6}\right|^{2} &
 \bar{a}_{6}\bar{a}_{7}^{\ast} &
 \bar{a}_{6}\bar{a}_{8}^{\ast} &
 \bar{a}_{6}\bar{a}_{9}^{\ast}\gamma \\ 
\bar{a}_{7}\bar{a}_{1}^{\ast}\gamma &
 \bar{a}_{7}\bar{a}_{2}^{\ast} &
 \bar{a}_{7}\bar{a}_{3}^{\ast} &
 \bar{a}_{7}\bar{a}_{4}^{\ast} &
 \bar{a}_{7}\bar{a}_{5}^{\ast}\gamma &
 \bar{a}_{7}\bar{a}_{6}^{\ast} &
 \left|\bar{a}_{7}\right|^{2} &
 \bar{a}_{7}\bar{a}_{8}^{\ast} &
 \bar{a}_{7}\bar{a}_{9}^{\ast}\gamma \\ 
\bar{a}_{8}\bar{a}_{1}^{\ast}\gamma &
 \bar{a}_{8}\bar{a}_{2}^{\ast} &
 \bar{a}_{8}\bar{a}_{3}^{\ast} &
 \bar{a}_{8}\bar{a}_{4}^{\ast} &
 \bar{a}_{8}\bar{a}_{5}^{\ast}\gamma &
 \bar{a}_{8}\bar{a}_{6}^{\ast} &
 \bar{a}_{8}\bar{a}_{7}^{\ast} &
 \left|\bar{a}_{8}\right|^{2} &
 \bar{a}_{8}\bar{a}_{9}^{\ast}\gamma \\ 
\bar{a}_{9}\bar{a}_{1}^{\ast}\gamma^{4} &
 \bar{a}_{9}\bar{a}_{2}^{\ast}\gamma &
 \bar{a}_{9}\bar{a}_{3}^{\ast}\gamma &
 \bar{a}_{9}\bar{a}_{4}^{\ast}\gamma &
 \bar{a}_{9}\bar{a}_{5}^{\ast} &
 \bar{a}_{9}\bar{a}_{6}^{\ast}\gamma &
 \bar{a}_{9}\bar{a}_{7}^{\ast}\gamma &
 \bar{a}_{9}\bar{a}_{8}^{\ast}\gamma &
 \left|\bar{a}_{9}\right|^{2} \\ 
\end{array}
\right). \
\end{eqnarray}
Note the form of the matrix, in which there are regions unaffected
by dephasing, with decoherence effects along the edges. We show in
the next section that this gives rise to a class of states that can
exist in these decoherence free subspaces, where entanglement is
also be explicitly characterized. We see in the above general matrix
that there exist two timescales of off-diagonal element decay, $
\tau_{\rm 2-dec,\mathcal{D}}^{G(1)} = 2\big({1\over\Gamma_{\rm
2}}\big) \ {\rm and} \ \tau_{\rm 2-dec,\mathcal{D}}^{G(2)} =
\big({1\over 2\Gamma_{\rm 2}}\big) . $

The reduced density matrices are given by the following matrices,
\begin{eqnarray}
\rho^{{\rm G},\mathcal D}_{{\rm A}}\left(t\right)=
{\rm Tr}_{\rm B}\rho^{{\rm G},\mathcal D}_{{\rm AB}}\left(t\right) =
\left(
\begin{array}{lll}
 \left|\bar{a}_{1}\right|^{2} +
 \left|\bar{a}_{2}\right|^{2} +
 \left|\bar{a}_{3}\right|^{2}
 &
 \bar{a}_{3}\bar{a}_{6}^{\ast} +
 \bar{a}_{1}\bar{a}_{4}^{\ast}\gamma +
 \bar{a}_{2}\bar{a}_{5}^{\ast}\gamma \ \ \ \
 &
 \bar{a}_{2}\bar{a}_{8}^{\ast} +
 \bar{a}_{1}\bar{a}_{7}^{\ast}\gamma +
 \bar{a}_{3}\bar{a}_{9}^{\ast}\gamma
 \\
 \bar{a}_{6}\bar{a}_{3}^{\ast} +
 \bar{a}_{4}\bar{a}_{1}^{\ast}\gamma +
 \bar{a}_{5}\bar{a}_{2}^{\ast}\gamma \ \ \ \
 &
 \left|\bar{a}_{4}\right|^{2} +
 \left|\bar{a}_{5}\right|^{2} +
 \left|\bar{a}_{6}\right|^{2}
 &
 \bar{a}_{4}\bar{a}_{7}^{\ast} +
 \bar{a}_{5}\bar{a}_{8}^{\ast}\gamma +
 \bar{a}_{6}\bar{a}_{9}^{\ast}\gamma
 \\
 \bar{a}_{8}\bar{a}_{2}^{\ast} +
 \bar{a}_{7}\bar{a}_{1}^{\ast}\gamma +
 \bar{a}_{9}\bar{a}_{3}^{\ast}\gamma \ \ \ \
 &
 \bar{a}_{7}\bar{a}_{4}^{\ast} +
 \bar{a}_{8}\bar{a}_{5}^{\ast}\gamma +
 \bar{a}_{9}\bar{a}_{6}^{\ast}\gamma \ \ \ \
 &
\left|\bar{a}_{7}\right|^{2} +
 \left|\bar{a}_{8}\right|^{2} +
 \left|\bar{a}_{9}\right|^{2}
 \\
\end{array}
\right) \\
\rho^{{\rm G},\mathcal D}_{{\rm B}}\left(t\right)=
{\rm Tr}_{\rm A}\rho^{{\rm G},\mathcal D}_{{\rm AB}}\left(t\right) =
\left(
\begin{array}{lll}
 \left|\bar{a}_{1}\right|^{2} +
 \left|\bar{a}_{4}\right|^{2} +
 \left|\bar{a}_{7}\right|^{2}
 &
 \bar{a}_{7}\bar{a}_{8}^{\ast} +
 \bar{a}_{1}\bar{a}_{2}^{\ast}\gamma +
 \bar{a}_{4}\bar{a}_{5}^{\ast}\gamma \ \ \ \
 &
 \bar{a}_{4}\bar{a}_{6}^{\ast} +
 \bar{a}_{1}\bar{a}_{3}^{\ast}\gamma +
 \bar{a}_{7}\bar{a}_{9}^{\ast}\gamma
 \\
 \bar{a}_{8}\bar{a}_{7}^{\ast} +
 \bar{a}_{2}\bar{a}_{1}^{\ast}\gamma +
 \bar{a}_{5}\bar{a}_{4}^{\ast}\gamma \ \ \ \
 &
 \left|\bar{a}_{2}\right|^{2} +
 \left|\bar{a}_{5}\right|^{2} +
 \left|\bar{a}_{8}\right|^{2}
 &
 \bar{a}_{2}\bar{a}_{3}^{\ast} +
 \bar{a}_{5}\bar{a}_{6}^{\ast}\gamma +
 \bar{a}_{8}\bar{a}_{9}^{\ast}\gamma
 \\
 \bar{a}_{6}\bar{a}_{4}^{\ast} +
 \bar{a}_{3}\bar{a}_{1}^{\ast}\gamma +
 \bar{a}_{9}\bar{a}_{7}^{\ast}\gamma \ \ \ \
 &
 \bar{a}_{3}\bar{a}_{2}^{\ast} +
 \bar{a}_{5}\bar{a}_{5}^{\ast}\gamma +
 \bar{a}_{9}\bar{a}_{8}^{\ast}\gamma \ \ \ \
 &
\left|\bar{a}_{3}\right|^{2} +
 \left|\bar{a}_{6}\right|^{2} +
 \left|\bar{a}_{9}\right|^{2}
 \\
\end{array}
\right).\
\end{eqnarray}
Note that the single-qutrit reduced density matrices do not become
entirely diagonal. The reason, as we stated above, is the existence
of decoherence free subspaces.  In this case, there exists only one
timescale in which a subset of off-diagonal elements decay, $
\tau_{\rm 2-dec,\mathcal{D}}^{G(1)} = 2\big({1\over\Gamma_{\rm
2}}\big). $

The expression for negativity for the general class under the
collective dephasing channel, is similar in character to the
negativity found in the previous subsection for the multi-local
dephasing channel given by Eq. 5.  Therefore, for clarity, we defer
our discussion of the behavior of the negativity in the case of the
collective dephasing channel until after treating specific states to
the next section.

\section{Specific Classes}
In order better to distinguish decoherence and disentanglement
behavior, let us now turn our attention to two specific classes of
states. In the standard-basis representation, the generic pure state
of a two-qutrit system is $\ket{\Psi} = \bar{a}_{1}\ket{1} +
\bar{a}_{2}\ket{2} + \bar{a}_{3}\ket{3} + \bar{a}_{4}\ket{4} +
\bar{a}_{5}\ket{5} + \bar{a}_{6}\ket{6} + \bar{a}_{7}\ket{7} +
\bar{a}_{8}\ket{8} + \bar{a}_{9}\ket{9}$, a normalized state vector
with $\bar{a}_{i} \in \mathbb{C}$.  The generic class of two-qutrit
pure states represented by $\ket{\Psi}$ contains two subclasses of
interest, distinguished by their coherence behavior in large
timescales under collective dephasing noise, that is, dephasing in
which each qutrit interacts with the same collective noise $b_{\rm
AB}\left(t\right)$. One class is seen to be {\it fragile},
whereas the other is {\it robust}. \\
\noindent (i) The \textit{fragile} class $\ket{\phi} =
\bar{a}_{1}\ket{1} + \bar{a}_{5}\ket{5} + \bar{a}_{9}\ket{9}$, in
which $\bar{a}_{1}, \bar{a}_{5}, {\rm and}\ \bar{a}_{9}$ may be
non-zero and all other terms $\bar{a}_{i}=0$, has the forms
\vfill\eject
\begin{eqnarray}
\ket{\phi_{1}} &=& \bar{a}_{1}\ket{1} + \bar{a}_{5}\ket{5}\ ,\\
\ket{\phi_{2}} &=& \bar{a}_{1}\ket{1} + \bar{a}_{9}\ket{9}\ ,\\
\ket{\phi_{3}} &=& \bar{a}_{5}\ket{5} + \bar{a}_{9}\ket{9}\ .
\end{eqnarray}
\noindent (ii) The \textit{robust} class
$\ket{\psi} =
\bar{a}_{2}\ket{2} + \bar{a}_{3}\ket{3} + \bar{a}_{4}\ket{4} +
\bar{a}_{6}\ket{6} + \bar{a}_{7}\ket{7} + \bar{a}_{8}\ket{8}$, in which
all $\bar{a}_{i}$ listed may be non-zero and
$\bar{a}_{1} = \bar{a}_{9} =0$, has the forms
\begin{eqnarray}
\ket{\psi_{1}} &=& \bar{a}_{2}\ket{2} + \bar{a}_{4}\ket{4}\ ,\\
\ket{\psi_{2}} &=& \bar{a}_{3}\ket{3} + \bar{a}_{7}\ket{7}\ ,\\
\ket{\psi_{3}} &=& \bar{a}_{6}\ket{6} + \bar{a}_{8}\ket{8}\ .
\end{eqnarray}
The specific forms of the two classes above constitute Bell-like
states in the Hilbert space of our two-qutrit system, similarly to
the qubit-pair-state classification given in \cite{YE03}.
\subsection{Fragile Class}
The initial density matrix representing the two-qutrit fragile class
is given by
\begin{eqnarray}
\rho_{{\rm AB}}^{\rm F}\left(0\right)=P(\ket{\phi}) \equiv
\left(
\begin{array}{ccccc}
 \left|\bar{a}_{1}\right|^{2} & \cdots & \bar{a}_{1}\bar{a}_{5}^{\ast} & \cdots & \bar{a}_{1}\bar{a}_{9}^{\ast} \\ 
 \vdots & \ddots & \vdots & \ddots & \vdots \\ 
 \bar{a}_{5}\bar{a}_{1}^{\ast} & \cdots & \left|\bar{a}_{5}\right|^{2} & \cdots & \bar{a}_{5}\bar{a}_{9}^{\ast} \\ 
 \vdots & \ddots & \vdots & \ddots & \vdots \\ 
 \bar{a}_{9}\bar{a}_{1}^{\ast} & \cdots & \bar{a}_{9}\bar{a}_{5}^{\ast} & \cdots & \left|\bar{a}_{9}\right|^{2} \\ 
\end{array}
\right),\
\end{eqnarray}
where the dots denote the remainder of the density matrix, which is
filled out by zero entries.  Let us turn our attention to the
behavior of this state when subjected to the multi-local dephasing
$\mathcal{EF}$ acting on qutrits A and B. Recall that the
multi-local dephasing channel captures the features of each of the
local dephasing channels fully, so for this analysis, we consider
the multi-local dephasing channel only, which provides information
about each of the individual dephasing channels if one simply turns
off one of the channels.

\subsubsection{Fragile Class: Multi-Local Dephasing Channel $\mathcal{EF}$}
The density matrix subjected to noise is
\begin{eqnarray}
\rho^{{\rm F},\mathcal{EF}}_{{\rm AB}}\left(t\right) =
\left(
\begin{array}{ccccc}
 \left|\bar{a}_{1}\right|^{2} & \cdots &
 \bar{a}_{1}\bar{a}_{5}^{\ast}\gamma_{\rm A}\gamma_{\rm B} & \cdots &
 \bar{a}_{1}\bar{a}_{9}^{\ast}\gamma_{\rm A}\gamma_{\rm B} \\ 
 \vdots & \ddots & \vdots & \ddots & \vdots \\ 
 \bar{a}_{5}\bar{a}_{1}^{\ast}\gamma_{\rm A}\gamma_{\rm B} & \cdots &
 \left|\bar{a}_{5}\right|^{2} & \cdots &
 \bar{a}_{5}\bar{a}_{9}^{\ast}\gamma_{\rm A}^{2}\gamma_{\rm B}^{2} \\ 
 \vdots & \ddots & \vdots & \ddots & \vdots \\ 
 \bar{a}_{9}\bar{a}_{1}^{\ast}\gamma_{\rm A}\gamma_{\rm B} & \cdots &
 \bar{a}_{9}\bar{a}_{5}^{\ast}\gamma_{\rm A}^{2}\gamma_{\rm B}^{2} &
 \cdots & \left|\bar{a}_{9}\right|^{2} \\ 
\end{array}
\right),\
\end{eqnarray}
which has two different decoherence times of
$
\tau_{{\rm 2-dec},\mathcal{EF}}^{\rm F} =
\big({1\over\Gamma_{\rm 1}}\big) \ {\rm and} \
\tau_{{\rm 2-dec},\mathcal{EF}}^{\rm F} =
\big({1\over2\Gamma_{\rm 1}}\big).
$
By comparison, the disentanglement timescales deriving from the
negativity,
$
\mathcal{N}\left(\rho^{\mathcal{EF}}_{{\rm AB}}\left(t\right)\right)
= \frac{\left\| \rho^{{\rm T}_{\rm A}} \right\|_{1} - 1}{2}
= \frac{\left\| \rho^{{\rm T}_{\rm B}} \right\|_{1} - 1}{2}
=
\left(
\left|\bar{a}_{1}\right|\left|\bar{a}_{5}\right| +
\left|\bar{a}_{1}\right|\left|\bar{a}_{9}\right|
\right)\gamma_{\rm A}\gamma_{\rm B} +
\left|\bar{a}_{5}\right|\left|\bar{a}_{9}\right|
\gamma_{\rm A}^{2}\gamma_{\rm B}^{2}
$
are
$
\tau_{{\rm dis},\mathcal{EF}}^{\rm F(1)} =
\big({1\over\Gamma_{1}}\big) \
{\rm and} \
\tau_{{\rm dis},\mathcal{EF}}^{\rm F(2)} =
\big({1\over2\Gamma_{1}}\big)
$.

The reduced density matrices of qutrit A and B are
\begin{eqnarray}
\rho_{\rm reduced}\left(t\right) =
{\rm Tr}_{\rm B}\rho^{{\rm F},\mathcal{EF}}_{{\rm AB}}\left(t\right) =
{\rm Tr}_{\rm A}\rho^{{\rm F},\mathcal{EF}}_{{\rm AB}}\left(t\right) =
\left(
\begin{array}{ccccccccc}
 \left|\bar{a}_{1}\right|^{2} & 0 & 0 \\ 
  0 & \left|\bar{a}_{5}\right|^{2} & 0 & \\ 
  0 & 0 & \left|\bar{a}_{9}\right|^{2} \\ 
\end{array}
\right)\ .
\end{eqnarray}
We see that they are both initially fully mixed, so that no further
decoherence is possible.

Comparing timescales we see that decoherence never proceeds faster than disentanglement,
$
\tau_{{\rm dis},\mathcal{EF}}^{\rm F} \le \tau_{\rm 2-dec,\mathcal{EF}}^{\rm F} \ {\rm and} \
\tau_{{\rm dis},\mathcal{EF}}^{\rm F} \le \tau_{\rm 1-dec,\mathcal{EF}}^{\rm F} \
$
\subsubsection{Fragile Class: Collective Dephasing Channel $\mathcal{D}$}
The two-qutrit density matrix, as given by
\begin{eqnarray}
\rho_{\rm AB}^{{\rm F},\mathcal D}\left(t\right) =
\left(
\begin{array}{ccccc}
 \left|\bar{a}_{1}\right|^{2} & \cdots & \bar{a}_{1}\bar{a}_{5}^{\ast}\gamma^{4} &
 \cdots & \bar{a}_{1}\bar{a}_{9}^{\ast}\gamma^{4} \\ 
 \vdots & \ddots & \vdots & \ddots & \vdots \\ 
 \bar{a}_{5}\bar{a}_{1}^{\ast}\gamma^{4} & \cdots & \left|\bar{a}_{5}\right|^{2} &
 \cdots & \bar{a}_{5}\bar{a}_{9}^{\ast} \\ 
 \vdots & \ddots & \vdots & \ddots & \vdots \\ 
 \bar{a}_{9}\bar{a}_{1}^{\ast}\gamma^{4} & \cdots & \bar{a}_{9}\bar{a}_{5}^{\ast} &
 \cdots & \left|\bar{a}_{9}\right|^{2} \\ 
\end{array}
\right),\
\end{eqnarray}
decays according to a single timescale, $ \tau_{{\rm
2-dec},\mathcal{D}}^{\rm F} = \big({1\over{2 \Gamma}}\big) $. The
disentanglement, characterized via the negativity, $
\mathcal{N}\left(\rho^{\mathcal{D}}_{{\rm AB}}\left(t\right)\right)
= \frac{\left\| \rho^{{\rm T}_{\rm A}} \right\|_{1} - 1}{2} =
\frac{\left\| \rho^{{\rm T}_{\rm B}} \right\|_{1} - 1}{2} =
\left|\bar{a}_{5}\right| \left|\bar{a}_{9}\right| + \left(
\left|\bar{a}_{1}\right|\left|\bar{a}_{5}\right| +
\left|\bar{a}_{1}\right|\left|\bar{a}_{9}\right| \right)\gamma^{4}.
$ Disentanglement proceeds at only a single timescale, $\tau_{{\rm
dis},\mathcal{D}}^{\rm F} = \big({1\over{2 \Gamma}}\big)$.

The single qutrit matrices are always
\begin{eqnarray}
\rho_{\rm reduced}\left(t\right) =
{\rm Tr}_{\rm B}\rho^{{\rm F},\mathcal{D}}_{{\rm AB}}\left(t\right) =
{\rm Tr}_{\rm A}\rho^{{\rm F},\mathcal{D}}_{{\rm AB}}\left(t\right) =
\left(
\begin{array}{ccccccccc}
 \left|\bar{a}_{1}\right|^{2} & 0 & 0 \\ 
  0 & \left|\bar{a}_{5}\right|^{2} & 0 & \\ 
  0 & 0 & \left|\bar{a}_{9}\right|^{2} \\ 
\end{array}
\right)\ .
\end{eqnarray}
Again, we see that the single qutrit states are both fully mixed, so
that no further decoherence is possible. Comparing timescales, we
see that decoherence never proceeds faster than disentanglement:
$\tau_{{\rm dis},\mathcal{D}}^{\rm F} \le \tau_{\rm
2-dec,\mathcal{D}}^{\rm F}. $

\subsection{Robust Class}
The two-qutrit density matrix for the robust class, under the
multi-local dephasing channel, is given by
\begin{eqnarray}
\rho^{\rm R}_{{\rm AB}}\left(0\right) =
\left(
\begin{array}{ccccccccc}
 0 & 0 & 0 & 0 & 0 & 0 & 0 & 0 & 0 \\ 
 0 &
 \left|\bar{a}_{2}\right|^{2} &
 \bar{a}_{2}\bar{a}_{3}^{\ast} &
 \bar{a}_{2}\bar{a}_{4}^{\ast} &
 0 &
 \bar{a}_{2}\bar{a}_{6}^{\ast} &
 \bar{a}_{2}\bar{a}_{7}^{\ast} &
 \bar{a}_{2}\bar{a}_{8}^{\ast} & 0 \\ 
 0 &
 \bar{a}_{3}\bar{a}_{2}^{\ast} &
 \left|\bar{a}_{3}\right|^{2} &
 \bar{a}_{3}\bar{a}_{4}^{\ast} &
 0 &
 \bar{a}_{3}\bar{a}_{6}^{\ast} &
 \bar{a}_{3}\bar{a}_{7}^{\ast} &
 \bar{a}_{3}\bar{a}_{8}^{\ast} & 0 \\ 
 0 &
 \bar{a}_{4}\bar{a}_{2}^{\ast} &
 \bar{a}_{4}\bar{a}_{3}^{\ast} &
 \left|\bar{a}_{4}\right|^{2} &
 0 &
 \bar{a}_{4}\bar{a}_{6}^{\ast} &
 \bar{a}_{4}\bar{a}_{7}^{\ast} &
 \bar{a}_{4}\bar{a}_{8}^{\ast} & 0 \\ 
 0 & 0 & 0 & 0 & 0 & 0 & 0 & 0 & 0 \\ 
 0 &
 \bar{a}_{6}\bar{a}_{2}^{\ast} &
 \bar{a}_{6}\bar{a}_{3}^{\ast} &
 \bar{a}_{6}\bar{a}_{4}^{\ast} &
 0 &
 \left|\bar{a}_{6}\right|^{2} &
 \bar{a}_{6}\bar{a}_{7}^{\ast} &
 \bar{a}_{6}\bar{a}_{8}^{\ast} & 0 \\ 
 0 &
 \bar{a}_{7}\bar{a}_{2}^{\ast} &
 \bar{a}_{7}\bar{a}_{3}^{\ast} &
 \bar{a}_{7}\bar{a}_{4}^{\ast} &
 0 &
 \bar{a}_{7}\bar{a}_{6}^{\ast} &
 \left|\bar{a}_{7}\right|^{2} &
 \bar{a}_{7}\bar{a}_{8}^{\ast} & 0 \\ 
 0 &
 \bar{a}_{8}\bar{a}_{2}^{\ast} &
 \bar{a}_{8}\bar{a}_{3}^{\ast} &
 \bar{a}_{8}\bar{a}_{4}^{\ast} &
 0 &
 \bar{a}_{8}\bar{a}_{6}^{\ast} &
 \bar{a}_{8}\bar{a}_{7}^{\ast} &
 \left|\bar{a}_{8}\right|^{2} & 0 \\ 
 0 & 0 & 0 & 0 & 0 & 0 & 0 & 0 & 0 \\ 
\end{array}
\right)\label{robustTimeZero}\ .
\end{eqnarray}
The rows and columns of zeros in this matrix are exactly the rows
and columns that are affected by the collective dephasing channel,
giving rise to the robust character of this density matrix. For
analysis in this section, we consider, without loss of generality,
the first form of this class $\ket{\psi_{1}} = \bar{a}_{2}\ket{2} +
\bar{a}_{4}\ket{4}$, which allows the decoherence and
disentanglement timescales to be seen clearly.  If we were to
include the entire subclass, we would have the same behavior as our
forthcoming analysis.

\subsubsection{Robust Class: Multi-Local Dephasing Channel $\mathcal{EF}$}
The time-evolved density matrix for the multi-local dephasing
channel acting on $\ket{\psi_{1}}$ is given by
\begin{eqnarray}
\rho^{{\rm R},\mathcal EF}_{{\rm AB}}\left(t\right) =
\left(
\begin{array}{ccccccccc}
 0 & 0 & 0 & 0 & 0 & 0 & 0 & 0 & 0 \\ 
 0 & \left|\bar{a}_{2}\right|^{2} &
 0 & \bar{a}_{2}\bar{a}_{4}^{\ast}\gamma_{\rm A}\gamma_{\rm B} &
 0 & 0 & 0 & 0 & 0 \\ 
 0 & 0 & 0 & 0 & 0 & 0 & 0 & 0 & 0\\ 
 0 & \bar{a}_{4}\bar{a}_{2}^{\ast}\gamma_{\rm A}\gamma_{\rm B} & 0 &
 \left|\bar{a}_{4}\right|^{2} & 0 & 0 & 0 & 0 & 0 \\ 
 0 & 0 & 0 & 0 & 0 & 0 & 0 & 0 & 0 \\ 
 0 & 0 & 0 & 0 & 0 & 0 & 0 & 0 & 0 \\ 
 0 & 0 & 0 & 0 & 0 & 0 & 0 & 0 & 0 \\ 
 0 & 0 & 0 & 0 & 0 & 0 & 0 & 0 & 0 \\ 
 0 & 0 & 0 & 0 & 0 & 0 & 0 & 0 & 0 \\ 
\end{array}
\right)\
\end{eqnarray}
We see that the off-diagonal elements decay according to the
timescale $ \tau_{{\rm 2-dec},\mathcal{EF}}^{\rm R} =
\big({1\over\Gamma_{\rm 1}}\big) $. The disentanglement,
characterized via the negativity, is given by,
$\mathcal{N}\left(\rho^{\mathcal{D}}_{{\rm AB}}\left(t\right)\right)
= \frac{\left\| \rho^{{\rm T}_{\rm A}} \right\|_{1} - 1}{2} =
\frac{\left\| \rho^{{\rm T}_{\rm B}} \right\|_{1} - 1}{2} =
\left|\bar{a}_{2}\right| \left|\bar{a}_{4}\right| \gamma_{\rm
A}\gamma_{\rm B}$. Disentanglement proceeds at the same single
timescale, $\tau_{{\rm dis},\mathcal{EF}}^{\rm R} =
\big({1\over{\Gamma_{\rm 1}}}\big)$. The reduced density matrices
are given by
\begin{eqnarray}
\rho^{{\rm R}, \mathcal EF}_{\rm A}\left(t\right) =
{\rm Tr}_{\rm B}\rho^{{\rm R},\mathcal{EF}}_{{\rm AB}}\left(t\right) =
\left(
\begin{array}{lll}
 \left|\bar{a}_{2}\right|^{2} & 0 & 0 \\
 0 & \left|\bar{a}_{4}\right|^{2} & 0 \\
 0 & 0 & 0\\ 
\end{array}
\right),   \\
\rho^{{\rm R},\mathcal{EF}}_{\rm B}\left(t\right) =
{\rm Tr}_{\rm A}\rho^{{\rm R},\mathcal{EF}}_{{\rm AB}}\left(t\right)\left(
\begin{array}{lll}
 \left|\bar{a}_{4}\right|^{2} & 0 & 0 \\
 0 & \left|\bar{a}_{2}\right|^{2} & 0 \\
 0 & 0 & 0\\ 
\end{array}
\right)\ ,
\end{eqnarray}
always fully decohered as in the case of the fragile states. Again,
we see that decoherence never proceeds faster than disentanglement,
so that clearly the inequality $ \tau_{{\rm dis},\mathcal{D}}^{\rm
R} \le \tau_{\rm 2-dec,\mathcal{D}}^{\rm R}$ is satisfied.

\subsubsection{Robust Class: Collective Dephasing Channel $\mathcal{D}$}
Here, $\rho\left(t\right) = \rho\left(0\right)$ as defined above in
Eq. \ref{robustTimeZero}. and the state neither decoheres nor
disentangles: the off-diagonal elements remain unreduced and the
negativity remains at the maximum value of 1 for all times. These
states are robust against collective dephasing noise.

\subsection{Conclusions}
We have shown for this model of two-qutrit systems under the chosen
basis-specific dephasing noise, whether local, multi-local or
collective in nature, that disentanglement proceeds at least as fast
as decoherence. This result accords with previous studies of similar
nature for qubit systems. Because the case of two-dimensions in
quantum mechanics is often a special one in quantum mechanics, for
example, in the cases of the foundational theorems of Gleason and
Kochen and Specker, it is always important to examine cases beyond
that of qubits for the possibility of different behavior from that
exhibited in the special case of qubits. Our results suggest that
the relation between basis-dependent dephasing decoherence and
disentanglement found here can be expected to hold for pairs of
initially entangled qu-$d$-it systems under analogous dephasing
noise for general values of $d$. It remains an open question whether
or not this phenomenon can be generalized to an arbitrary
\emph{multi-partite} quantum systems, each of arbitrary dimension,
however, not least of all because good entanglement measures for
mixed states of such systems is lacking.



\begin{thebibliography}{12}

\bibitem{YE02} T. Yu and J. H. Eberly, Phys. Rev. B
{\bf 66}, 193306 (2002).

\bibitem{YE03} T. Yu and J. H. Eberly, Phys. Rev. B {\bf 68}, 165322
(2003).

\bibitem{YE04} T. Yu and J. H. Eberly, Phys. Rev. Lett. {\bf
93}, 140404 (2004).

\bibitem{YE06}
T. Yu and J. H. Eberly, Opt. Commun. {\bf 264}, 393 (2006).

\bibitem{TP05} D. Tolkunov and V. Privman, P. K. Aravind, Phys. Rev. A {\bf 71},
060308(R) (2005).

\bibitem{STP06} D. Solenov, D. Tolkunov, and V.
Privman, Phys. Lett. A {\bf 359}, 81 (2006).

\bibitem{AY07} K. Ann and G. S. Jaeger, Phys. Rev. B, in press.



\bibitem{YE03} T. Yu and J. H. Eberly, Phys. Rev. B
{\bf 68}, 165322 (2003).

\bibitem{FS06} Z. Ficek and S. Swain, Phys. Rev. A {\bf 69}, 023401 (2004).

\bibitem{DJ06} L. Derkacz and L. Jakobczyk, Phys. Rev. A {\bf 74}, 03213 (2006).

\bibitem{CM00} C.M. Caves and G.J. Milburn, Opt. Commun. {\bf 179}, 439 (2000).

\bibitem{CW06} A. Checinska and K. Wodkiewicz, quant-ph/0610127(v2) (2006).

\bibitem{VW02} G. Vidal and R.F. Werner, Phys. Rev. A {\bf 65}, 032314 (2002).

\bibitem{H97} P. Horodecki, Phys. Lett. A {\bf 232}, 333 (1997).

\end{thebibliography}
\end{document}